# Observation of quantum friction in solid $^4He$

Almog Danzig, Ori Scaly and Emil Polturak

*Address:* Department of Physics, Technion - Israel Institute of Technology, Haifa 32000, Israel

*Electronic Address:* emilp@physics.technion.ac.il

## Abstract

Classical sliding friction is dominated by the slip-stick mechanism, where contacts between two bodies are alternately formed and sheared as the bodies move past each other[1–3]. When the interface between two bodies is perfectly smooth, classical friction goes to zero[4], a state called superlubricity[5,6]. In this limit, much weaker mechanisms, called quantum friction[7–9] are predicted. These mechanisms are based on an exchange of elementary excitations between two bodies moving relatively to each other. For the friction to be called "quantum", the excitations must arise from the quantum mechanical description of the bodies. Photons and phonons are such excitations. Friction results from an irreversible momentum transfer from these excitations to the bodies, affecting their motion. We measured the friction force between crystallites of solid $^4He$ moving relative to each other at low temperatures. Our data are in excellent agreement with the concept of quantum phonon friction[8].

"Phonon friction" is a term used to describe energy transfer from interface deformations to phonons[10–12] which converts elastic energy to heat. With a smooth interface, this friction is absent. Quantum phonon friction as proposed by Popov[8] deals with phonons in two solids separated by a perfect interface, moving with respect to each other. Phonons are transmitted back and forth across this interface. At rest, the phonon populations in the two solids are identical. With a relative velocity, transmitted phonons are Doppler shifted. Phonons have an effective viscosity, resulting from momentum transfer to the crystal via Ümklapp processes. The relative velocity creates an imbalance of these Ümklapp processes between phonons having momenta parallel and anti-parallel to the velocity. This causes a net transfer of momentum to the center of mass of the solids, opposing the motion. The rate of momentum transfer is the friction force. At low temperatures, the predicted stress is $\sigma = V[\pi^2(kT)^4 \exp(-\theta_D/2T)]/[45\hbar^3 c_\perp^4]$, proportional to the velocity $V$, the density of phonons $(kT)^4$, and the probability of an Ümklapp process, $\exp(-\theta_D/2T)$, with $\theta_D$ the Debye temperature. The irreversible nature of the Ümklapp process is in line with the Leggett-Caldeira scenario of dissipation in a quantum system[13].

In a real system, phonon friction and classical friction will both be present. To detect phonon friction, classical friction should be minimized. One way is to use a soft cantilever oscillating close to the surface of the sample[14] and eliminate the contact between the bodies. Changes in the electronic friction associated with the superconducting transition[15] were demonstrated in this way[12]. Unfortunately, the non-contact approach is inapplicable to solid $^4He$ which exists only under pressure, so our experiment is conducted on a solid contained inside a pressurized cell (supplementary information). Our cell has an annular geometry, forming a part of a torsional oscillator (TO) shown in Fig. 1 as an inset. The TO was introduced to Helium physics by J. Reppy[16]. Its resonant frequency $\omega = \sqrt{\kappa/I}$, where $\kappa$ is the torsion constant and $I$ the moment of inertia. In the experiment, we grow a single hcp $^4He$ crystal inside the cell at a constant temperature and solidification pressure. It takes more than 5 days to grow a single crystal of 1 $cm^3$ size. Once the cell is full of solid, we cool it by $0.1K$. In our geometry, the central post of the annulus (Fig. 1)) prevents a uniform thermal contraction of the crystal. Consequently, the crystal is strained beyond its critical strain ($\sim 3 \times 10^{-8}$) and decomposes into many crystallites[17] (supplementary information). In that state, the resonant frequency of the TO spontaneously increases. In our temperature regime both the torsion constant $\kappa$ and the amount of solid inside the cell are fixed and independent of temperature. The resonant frequency can therefore increase only if the effective moment of inertia of the TO decreases. This can happen if part of the crystallites inside the cell become decoupled from the wall and do not move with the TO as a rigid body. In different cells, the part decoupled from the walls was between 10% and 30% of the moment of inertia of the solid $^4He$ in the cell. The rest (70% − 90%) of the solid crystallites move with walls of the TO. In the reference frame of the TO, the coupled part of the solid is static while the decoupled part of the solid moves, so there should be relative motion between these two parts. The oscillation amplitude is in the 10-100 nm range, so the relative motion is small.

To confirm that such a relative motion exists inside the cell, we developed an acoustic AFM-like sensor operating in-situ[18]. At the microscopic level, relative motion of two crystallites along the interface separating them generates vibrations resulting from the periodic atomic structure of the interface. With a speed $V$, these vibrations have a frequency $f = V/d$, where $d$ is the interatomic spacing. In absence of a relative motion or if the motion is random, such vibrations would not exist or average to zero over time. Indeed, we detected no vibrations when the cell contained liquid $^4He$ (random atomic motion) or a single crystal (single crystal is rigid, no relative motion). Vibrations were detected only when the cell contained a decomposed crystal[18]. The detection of these vibrations confirmed the existence of relative motion inside the cell.

In the experiment, we measured the resonant amplitude and the resonant frequency of the TO as a function of the temperature and the driving torque. Dissipation resulting from friction inside the cell lowers the amplitude. The minimal dissipation we can detect is in the $10^{-13} Watt$ range. The equation of motion of the TO is:

**Equation 1:**

$$I\ddot{\theta}(t) + \left(\gamma_{bulk} + \gamma_{interface}(1 - \beta e^{-i\alpha})\right)\dot{\theta}(t) + \kappa\theta(t) = \tau_0 e^{i\omega t}$$

Here $\theta(t)$ is the angular displacement of the TO, $I$ is the moment of inertia of the TO and the solid $^4He$ moving with it, and $\tau_0$ is the external driving torque. Since the dissipation is small, we can write the damping term as a linear combination of two friction mechanisms. The first coefficient, $\gamma_{bulk}$, represents the internal friction of the TO and the Helium mass coupled to the TO. The second coefficient, $\gamma_{interface}$, represents interfacial friction between the coupled and the decoupled solid $^4He$. We write the angular velocity of the decoupled solid $^4He$ as $Re[\dot{\theta}(t)\beta e^{-i\alpha}]$ with $0 \leq \beta \leq 1$. This means that the decoupled helium moves with a smaller amplitude than the wall by a factor $\beta$ and its motion has some phase shift $\alpha$ relative to the motion of the wall. The solution of the equation of motion (supplementary information) yields the amplitude and frequency of the TO at resonance as a function of the friction coefficients $\gamma_{bulk}$ and $\gamma_{interface}$. The temperature dependence of $\gamma_{bulk}$ was determined separately by measuring the amplitude vs. temperature of solid $^4He$ samples which showed no mass decoupling. In this case there are no interfaces and hence no interfacial friction. We found that $\gamma_{bulk}$ is independent of the temperature. Consequently, all the temperature dependence of the amplitude comes from interfacial friction $\gamma_{interface}$. In a previous experiment[19] we found that in the range $0.5K < T < 1.6K$, interfacial friction results from climb of individual dislocations. In this case, the friction coefficient $\gamma_{interface}$ is proportional to $exp(-\frac{E_J}{T})/T$, where $E_J$ is the energy of a jog[20]. This mechanism can arise if the interfaces are slightly imperfect. On the scale of the relative motion (*nm* range), imperfections can be small enough to be relieved by climb of individual dislocations.

During our previous experiment[19] we noticed that for $T \geq 1.6K$, the dissipation started to increase much faster, indicating that another friction mechanism becomes important. To grow $^4He$ crystals which remain solid at higher temperatures, we built a TO which can withstand much higher pressures (solidification pressure of $^4He$ increases with temperature, see supplementary information). With this TO, we grew crystals with a melting temperature of *2.5K* (solidification pressure of *59 bar*) and did measurements over the range *0.5K–2.4K*. The additional friction mechanism we added to our model is phonon friction[8]. The interfacial friction coefficient now includes two terms: $\gamma_{interface} = Aexp(-\frac{E_J}{T})/T + BT^4 exp(-\frac{\theta_D}{2T})$. The first term is due to climb of dislocations and the second to phonon friction. Here, $A$ and $B$ are constants. The equations for the resonant amplitude and frequency of the TO are given in supplementary information. Each friction mechanism has a different temperature dependence. The contribution of each term to the total friction coefficient is shown in Figure 1 below.

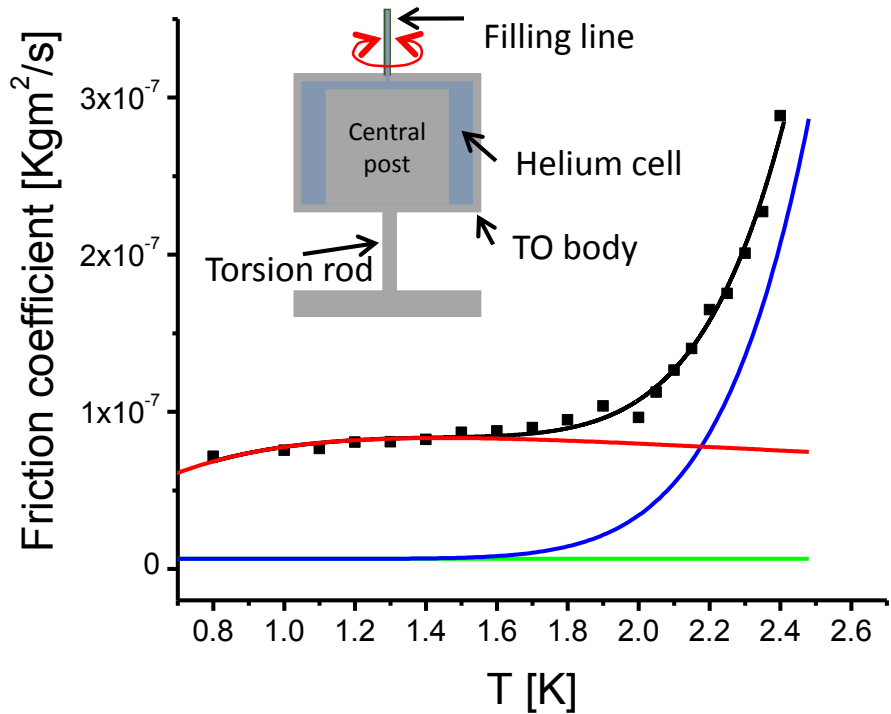

**Figure 1**: Temperature dependence of various mechanisms of friction. Green line- bulk friction, red line- dislocation climb, blue line- phonon friction, black line- total friction coefficient. The weight of each mechanism was determined by fitting experimental data (black symbols). The units are due to Eq. 1 describing a torque. **Inset:** Schematic cross section of the Torsional Oscillator (TO) containing an annular cell filled with solid $^4He$. The TO is made from *Be-Cu*. The annulus has a radius of *6.5mm*, width of *2mm*, and *10mm* height.

The amplitude of the TO and the fit to our model are both shown in Figure 2 for a range of driving torques spanning an order of magnitude. A similar fit of the resonant frequency is shown in the lower panel of Figure 3. Obviously, the fits are excellent. According to Popov's model[8], phonon friction depends linearly on the velocity. This assumption implies that the scale parameter *B* of the phonon friction should be independent of the driving force (supplementary information). At the top panel of Figure 3 we plot *B* vs. the driving force. Within the error bars, *B* is a constant, confirming the linear dependence of phonon friction on velocity.

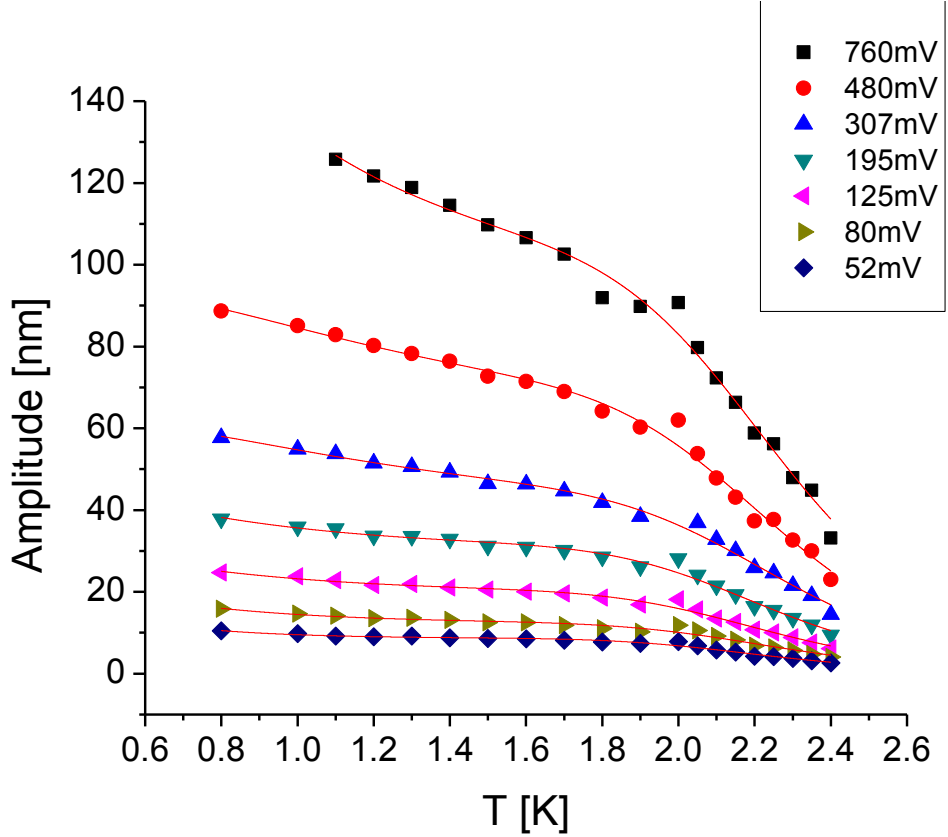

Figure 2: Resonant oscillation amplitude of the TO vs. temperature at different driving torques. The solid lines are fits to the model (see supplementary information). The amplitude in absolute units was determined using optical interferometry.

Since our interpretation involves both dislocations and phonons, we checked whether an interaction between them contributes to the friction[21]. One such interaction, "phonon wind", describes scattering of phonons by moving dislocations. The friction coefficient is proportional to $T^5$ when $T \ll \theta_D$. Another mechanism, called the flutter effect, involves excitation of vibrations of pinned dislocations by phonons. The flutter friction coefficient is proportional to $T^3$ when $T \ll \theta_D$. We tried to fit our data using these models (supplementary material). These fits did not converge well and the fitting parameters had very large error bars. In contrast, the fit to Popov's model converged very well.

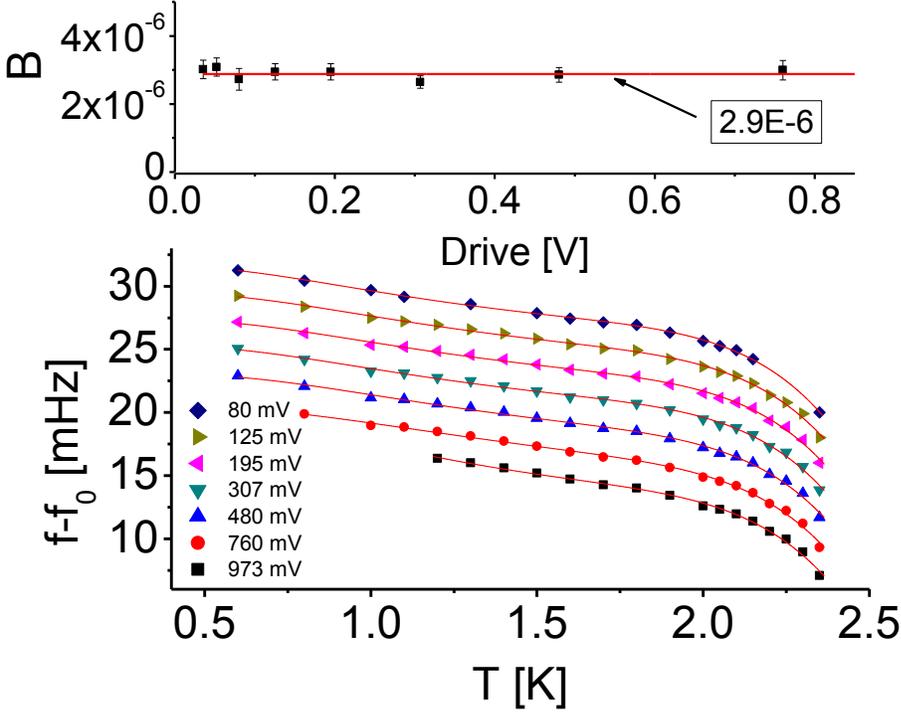

**Figure 3: Top panel:** $B$, the scale parameter of the phonon friction vs. drive. Red line is a fit to a constant (see supplementary information). **Bottom panel:** Variation of the resonant frequency vs. temperature with different driving torques. Here, $f_0 = 445.8 Hz$. For clarity, the data sets are offset vertically by *2 mHz*. Solid lines are fits to our model (see supplementary information).

Before concluding, we consider the magnitude of the friction forces in this work. If we take it that the frictional force is uniform over the annular channel, then the measured phonon frictional stress is ≈ 30 times larger than predicted[8]. This may perhaps be explained by recalling that phonons undergoing an Ümklapp process have wavevectors close to the zone boundary. In this region, the phonon dispersion relation $\omega(k)$ is flat, and so the phonon density of states should be enhanced as $(\nabla_k \omega)^{-1}$. Popov[8] uses the Debye model which does not take this into account. Neutron scattering data on solid $^4He$[22] show an enhancement of the density of states of this magnitude. Therefore, an explanation in these terms may be possible. Second, we compare the classical part of the friction shear stress which we measured to what is found with usual materials. A typical range for our system is ≈ $10^{-4} Pa$. To the best of our knowledge, the lowest friction stress reported to date was $5 \cdot 10^4 Pa$ measured between two incommensurate graphite surfaces[23] at room temperature. Evidently, classical friction associated with solid $^4He$ is orders of magnitude smaller, which we believe is the reason why a quantum mechanism of friction could be detected.

### *Acknowledgments*

We thank S. Hoida and L. Yumin for assistance and Dr. Lev Melnikovsky for fruitful discussions. This work was supported by the Israel Science Foundation and by the Technion Fund for Research.

# Supplementary information

## Crystal Growth

For this experiment, we grow single crystals of hcp solid helium at a constant temperature of 2.5K and a constant pressure of 59 bar, marked by the cross in the phase diagram (adjacent figure). The crystal is grown from the fluid phase. The procedure of growing single crystals of Helium was developed during previous neutron scattering studies. The limiting factor on the growth rate is the removal of the latent heat during solidification.

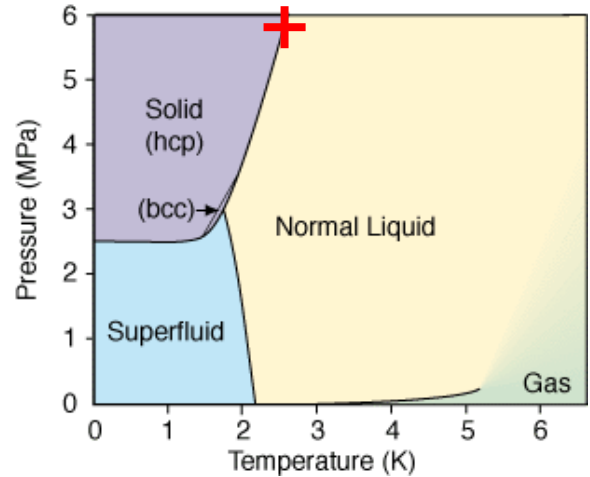

## Disordering a single crystal:

In the adjacent figure we show a typical Bragg diffraction spectrum of a He crystal measured by elastic neutron scattering[1,2]. With a single crystal in the cell, intense Bragg peaks are seen, such as the one shown by blue symbols. The angular width of these peaks is determined by the resolution of the spectrometer. The red symbols show the Bragg diffraction spectrum of the same solid after cooling the cell by a few tens of millikelvin. These (red) Bragg peaks indicate that the single crystal has now disintegrated into an array of a crystallites with a range of crystallographic orientations, all of them shifted from the orientation of the parent single crystal. In our work, we follow exactly the same procedure in order to disorder the He crystals inside the TO.

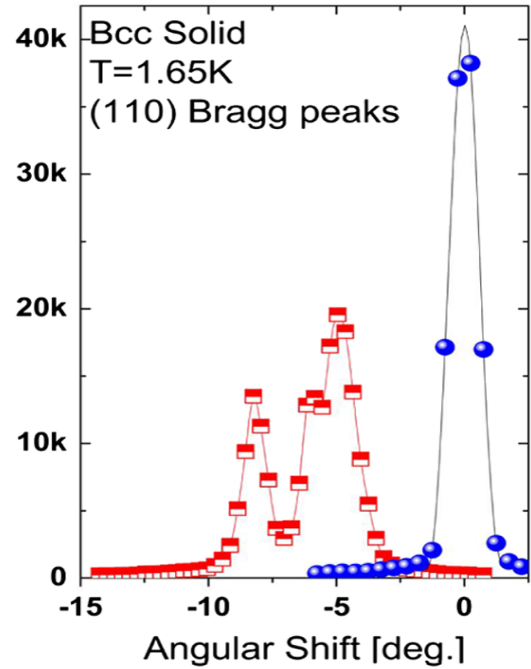

## Data analysis

Here, we review the data analysis in detail. The equation of motion of the TO is:

**Equation 2**

$$I\ddot{\theta}(t) + \left(\gamma_{bulk} + \gamma_{interface}(1 - \beta e^{-i\alpha})\right)\dot{\theta}(t) + \kappa\theta(t) = \tau_0 e^{i\omega t}$$

Here $\theta(t)$ is the angular displacement, *I* is the moment of inertia of the TO and the solid $^4$He moving with it, $\kappa$ is the torsion constant and $\tau_0$ is the external driving torque. In our model, where the "decoupled mass" is coupled to the TO only by a friction force, one can show that $\cos(\alpha) = \beta$. The quality factor *Q* of the TO is high (in excess of $10^5$), meaning that the dissipation is very small. We can therefore write the total dissipation as a linear combination of two friction mechanisms. Both mechanisms are linearly proportional to the angular speed $\dot{\theta}$. The first part, denoted by $\gamma_{bulk}\,\dot{\theta}$, describes the internal dissipation of the TO body and the bulk solid helium rigidly attached to the walls. The second part, $\gamma_{interface}\,\dot{\theta}$, represents the interfacial friction between the coupled and the decoupled solid during their relative motion.

During the experiment, we use a phase locked loop to keep the TO oscillating at its resonance frequency. Substituting a periodic solution for $\theta(t)$ in Equation 2, we get the resonant amplitude and frequency:

**Equation 3**

$$\theta_0 = \frac{\tau_0}{\omega\gamma_{bulk} + \omega\gamma_{interface}(1-\beta^2)}$$

**Equation 4**

$$\omega = \sqrt{\frac{\kappa}{I}} - \frac{\gamma_{interface}\beta\sqrt{1-\beta^2}}{2I}$$

## Fitting data

As explained in the manuscript, we take the interfacial friction coefficient to have the form: $\gamma_{interface} = A\exp(-\frac{E_J}{T})/T + BT^4\exp(-\frac{\theta_D}{2T})$ where the first term describes friction through climb of dislocation and the second is the phonon friction term. Here, *A* and *B* are constant coefficients. We also recall that $\gamma_{bulk}$ is a constant, independent of temperature. With this parametrization, Equation 3 and Equation 4 can be rewritten in a form which is convenient in fitting data. First, we write the resonant amplitude:

**Equation 5**

$$Amplitude[V] = \frac{1}{\tilde{E} + \tilde{A}\dfrac{\exp(-E_J/T)}{T} + \tilde{B}T^4\exp(-\dfrac{\theta_D}{2T})}$$

The temperature dependence of the amplitude is written explicitly in **Equation 5**, so that the fitting parameters $\tilde{E}, \tilde{A}, \tilde{B}$ are independent of temperature. Experimental values of the amplitude and driving force (Drive) are in Voltage units. To obtain the friction coefficient in physical units, we use a calibration factor which we measured for our experimental system, denoted by $c_3$:

**Equation 6**

$$\tilde{E} = \frac{\gamma_{bulk}}{c_3 I \cdot Drive}$$

$$\tilde{A} = A \frac{(1-\beta^2)}{c_3 I \cdot Drive}$$

$$\tilde{B} = B \frac{(1-\beta^2)}{c_3 I \cdot Drive}$$

The factor $\beta$ defined in Equation 2 is in the range $0 < \beta < 1$. The factor $(1-\beta^2)$ is a constant which we estimate as $0.6 < (1-\beta^2) < 1$. Using the parameters determined by the fit, we calculate the experimental friction coefficient shown in Figure 1 of the manuscript:

**Equation 7**

$$\gamma_{bulk} + (1-\beta^2)\gamma_{interface} = \frac{c_3 I \cdot Drive}{Amplitude}$$

Note that because we describe our system with a torque equation, the units of the friction coefficient $\gamma$ are $\frac{Kg \cdot m^2}{s}$.

In order to make a quantitative comparison with the phonon friction model, we are actually interested in the magnitude of $B$, the scale factor of the phonon term in the friction coefficient. In Equation 6, the parameter $\tilde{B} \propto \frac{B}{Drive}$. Multiplying the values of $\tilde{B}$ taken from the fit by the drive and by the different constants, we get $B$. According to the model, this scale factor $B$ should be a constant, independent of the drive. In Fig. 3 of the manuscript we plot the values of B vs. the driving voltage, confirming that B is indeed a constant.

We can write a relation similar to Equation 5 for the resonant frequency:

**Equation 8**

$$frequency = f_0 - C \frac{exp(-E_J/T)}{T} - DT^4 exp(-\frac{\theta_D}{2T})$$

Where the fitting parameters are:

**Equation 9**

$$C = \frac{\beta\sqrt{1-\beta^2}}{4\pi I} A$$

$$D = \frac{\beta\sqrt{1-\beta^2}}{4\pi I}B$$

and $f_0$ is the resonant frequency measured at our lowest temperature. The data and the fit are shown in Fig. 3 of the manuscript.

Beside the temperature dependence of the friction coefficient, we can compare the theoretical and experimental magnitude of *B*. We assume that the interface area between the coupled and decoupled helium is the surface area of the annular sample space. The ratio between measured and calculated *B* is approximately *30*. We believe that the uncertainty of the interfacial area used in this calculation can be a factor of order unity, but that still leaves an order of magnitude to be explained. One possible explanation for this difference are discussed in the last paragraph of the manuscript.

## Other theories of interaction between phonons and moving dislocation

In the figure below we show the friction coefficient vs. temperature, obtained by fitting data to Eq. 4. For the phonon term in this equation, we alternately used the Popov mechanism, the "phonon wind"[3], and the "flutter effect"[4]. Only Popov's model fits the experimental data well. Moreover, fits of the data done using the "phonon wind" and flutter effect did not converge well and the fitting parameters had very large error bars. In contrast, the fit to Popov's model converged very well.

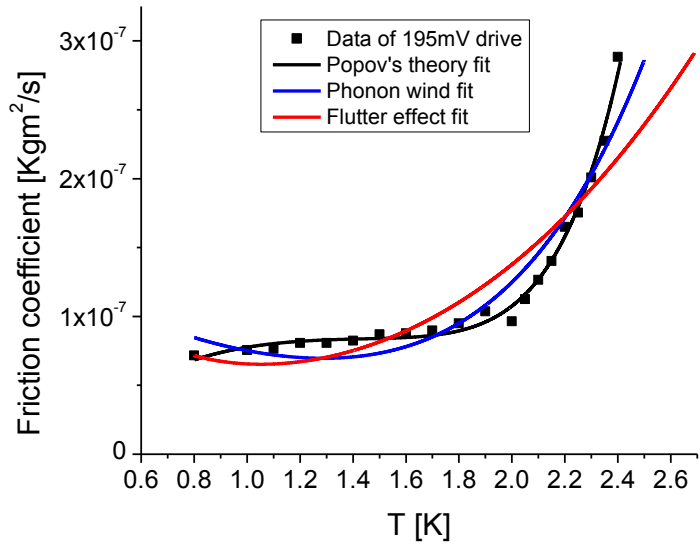

Figure 4: Friction coefficient vs. temperature data at a driving force of **$195 mV$** (black symbols). Blue and red lines are fits using friction mechanisms associated with interaction between phonons and moving dislocations[3,4].

### References:

1. Pelleg, O. *et al.* Observation of macroscopic structural fluctuations in bcc solid $^4$He. *Phys. Rev. B* **73,**

   024301 (2006).